\documentclass[onecolumn,prl,showpacs,preprintnumbers,amsmath,amssymb]{revtex4}
\usepackage{graphicx,epsfig}
\usepackage{dcolumn}
\usepackage{bm}
\begin{document}
\title{How glassy are orientational dynamics of rodlike molecules near the
isotropic-nematic transition?}
\author{Biman Jana, Dwaipayan Chakrabarti, and Biman Bagchi}
\email{bbagchi@sscu.iisc.ernet.in}
\affiliation{Solid State and Structural Chemistry Unit, Indian Institute of
Science, Bangalore 560012, India}

\begin{abstract}

In an attempt to quantitatively characterize the recently observed slow 
dynamics in the isotropic and nematic phase of liquid crystals, 
we investigate the single-particle orientational dynamics of rodlike
molecules across the isotropic-nematic transition in computer
simulations of a family of model systems of thermotropic liquid
crystals. Several remarkable features of glassy dynamics are on display
including non-exponential relaxation, dynamical heterogeneity, and
non-Arrhenius temperature dependence of the orientational relaxation
time. In order to obtain a {\it quantitative} measure of glassy
dynamics in line with the estbalished methods in supercooled liquids,
 we construct a relaxation time versus scaled inverse
temperature plot, and demonstrate  that one can indeed define a 'fragility
index' for thermotropic liquid crystals, that depends on density and
aspect ratio. The values of the fragility parameter are surprisingly
in the range one observed for glass forming liquids. A plausible
correlation between the energy landscape features and the observed
fragility is discussed.
\end{abstract}


\maketitle

\section{introduction}

Thermotropic liquid crystals exhibit exotic phase behavior upon temperature
variation. The nematic phase is rich with a long-ranged orientational order but
lacks translational order. The isotropic-nematic (I-N) phase transition, which is
believed to be weakly first order in nature with certain characteristics of the
continuous transition, has been a subject of immense attention in condensed matter
physics and material sciences \cite{deGennes-book-1993,Chandrasekhar-book-1992}.
In contrast, the dynamics of thermotropic liquid crystals have been much less
studied, the focus being mostly on the long-time behavior of orientational
relaxation near the I-N transition \cite{deGennes-book-1993}. A series of optical
Kerr effect (OKE) measurements have, however, recently studied collective
orientational relaxation in the isotropic phase near the I-N transition over a
wide range of time scales \cite{Gottke-JCP-2002}. The dynamics have been found to
be surprisingly rich, the most intriguing feature being the power law decay of the
OKE signal at short-to-intermediate times \cite{Gottke-JCP-2002}. The relaxation
scenario appears to be strikingly similar to that of supercooled molecular liquids
\cite{Cang-JCP-2003}, even though the latter do not undergo any thermodynamic
phase transition. Although the analogous dynamics have been investigated in
subsequent studies \cite{Jose-PRE-2005,Chakrabarty-PRE-2006}, a quantitative
estimation of glassy dynamics of rodlike molecules near the I-N transition still
eludes us.

The prime objective of this paper is to provide a quantitative measure of glassy
dynamics near the I-N transition. To this end, we have undertaken molecular
dynamics simulations of a family of model systems consisting of rodlike molecules
across the I-N transition in search of glassy behavior. Given the involvement of
the phase transition to an orientationally ordered mesophase upon lowering the
temperature, we choose to probe the single-particle orientational dynamics. We
have defined a 'fragility index' and explored plausible correlation of the
features of the underlying energy landscape with the observed fragility in analogy
with supercooled liquids.

\section{Model and details of the simulation}

The systems we have studied consist of ellipsoids of revolution. The Gay-Berne (GB)
pair potential \cite{Gay-JCP-1981}, that is well established to serve as a model potential
for systems of thermotropic liquid crystals, has been employed. The GB pair potential,
which uses a single-site representation for each ellipsoid of revolution, is an elegant
generalization of the extensively used isotropic Lennard-Jones potential to incorporate
anisotropy in both the attractive and the repulsive parts of the interaction \cite{Gay-JCP-1981,
Bates-JCP-1999}.In the GB pair potential, $i$th ellipsoid of revolution is represented by the position
${\bf r}_{i}$ of its center of mass and a unit vector ${\bf e}_{i}$ along the
long axis of the ellipsiod. The interaction potential between two ellipsoids of
revolution $i$ and $j$ is given by

\begin{equation}
U_{ij}^{GB}({\bf r}_{ij},{\bf e}_{i},{\bf e}_{j})  =
4\epsilon({\bf \hat r}_{ij},{\bf e}_{i},{\bf e}_{j})(\rho_{ij}^{-12} -
\rho_{ij}^{-6})
\end{equation}
where
\begin{equation}
\rho_{ij} = \frac{r_{ij} - \sigma({\bf \hat r}_{ij},{\bf e}_{i},{\bf e}_{j})
+ \sigma_{ss}}{\sigma_{ss}}.
\end{equation}
Here $\sigma_{ss}$ defines the thickness or equivalently, the separation between
the two ellipsoids of revolution in a side-by-side configuration, $r_{ij}$ is the distance between the
centers of mass of the ellipsoids of revolution $i$ and $j$, and
${\bf \hat r}_{ij} = {\bf r}_{ij} / r_{ij} $ is a unit vector along the
intermolecular separation vector ${\bf r}_{ij}$. The molecular shape
parameter $\sigma$ and the energy parameter $\epsilon$ both depend on the
unit vectors ${\bf e}_{i}$ and ${\bf e}_{j}$ as well as on
${\bf \hat r}_{ij}$ as given by the following set of equations:
\begin{equation}
\sigma({\bf \hat r}_{ij},{\bf e}_{i},{\bf e}_{j}) = \sigma_{0}\left[1 -
\frac{\chi}{2} \left\{\frac{({\bf e}_{i}\cdot{\bf \hat r}_{ij} +
{\bf e}_{j}\cdot{\bf \hat r}_{ij} )^{2}}
{1 + \chi({\bf e}_{i}\cdot{\bf e}_{j})} +
\frac{({\bf e}_{i}\cdot{\bf \hat r}_{ij} -
{\bf e}_{j}\cdot{\bf \hat r}_{ij})^{2}}{1 - \chi({\bf e}_{i} \cdot
{\bf e}_{j})}\right\}\right]^{-1/2}
\end{equation}
with $\chi = (\kappa^{2} - 1) / (\kappa^{2} + 1)$ and
\begin{equation}
\epsilon({\bf \hat r}_{ij},{\bf e}_{i},{\bf e}_{j}) = \epsilon_{0}
[\epsilon_{1}({\bf e}_{i},{\bf e}_{j})]^{\nu}
[\epsilon_{2}({\bf \hat r}_{ij},{\bf e}_{i},{\bf e}_{j})]^{\mu}
\end{equation}
where the exponents $\mu$ and $\nu$ are adjustable parameter, and
\begin{equation}
\epsilon_{1}({\bf e}_{i},{\bf e}_{j}) =
[1 - \chi^{2}({\bf e}_{i}\cdot{\bf e}_{j})^{2}]^{-1/2}
\end{equation}
and
\begin{equation}
\epsilon_{2}({\bf \hat r}_{ij},{\bf e}_{i},{\bf e}_{j}) = 1 -
\frac{\chi ^{\prime}}{2} \left[\frac{({\bf e}_{i}\cdot{\bf \hat r}_{ij} +
{\bf e}_{j}\cdot{\bf \hat r}_{ij} )^{2}}
{1 + \chi^{\prime}({\bf e}_{i}\cdot{\bf e}_{j})} +
\frac{({\bf e}_{i}\cdot{\bf \hat r}_{ij} -
{\bf e}_{j}\cdot{\bf \hat r}_{ij})^{2}}{1 - \chi ^{\prime}({\bf e}_{i} \cdot
{\bf e}_{j})}\right]
\end{equation}
with $\chi^{\prime} = (\kappa^{\prime ~ 1/\mu} - 1) /
(\kappa^{\prime ~ 1/\mu} + 1)$. Here
$\kappa = \sigma_{ee}/\sigma_{ss}$ is the aspect ratio of the ellipsoid of
revolution with $\sigma_{ee}$ denoting the separation between two ellipsoids of
revolution in a end-to-end configuration, and $\sigma_{ss} = \sigma_{0}$, and
$\kappa^{\prime} = \epsilon_{ss}/\epsilon_{ee}$, where $\epsilon_{ss}$ is the
depth of the minimum of the potential for a pair of ellipsoids of revolution aligned
in a side-by-side configuration, and $\epsilon_{ee}$ is the corresponding depth for the end-to-end
alignment. $\epsilon_{0}$ is the depth of the minimum of the pair potential between two ellipsoids 
of revolution alligned in cross configuration. The GB pair potential defines a
family of models, each member of which is characterized by the values chosen for the set of
four parameters $\kappa, \kappa^{\prime}, \mu,$ and $\nu$, and is represented by GB($\kappa,
\kappa^{\prime}, \mu, \nu$) \cite{Bates-JCP-1999}. Three systems, namely GB(3, 5, 2, 1), GB(3.4, 5, 2, 1),
and GB(3.8, 5, 2, 1), that differ in the aspect ratio have been investigated. Molecular dynamics
simulations have been performed with each of these systems, consisting of 500 ellipsoids
of revolution, in a cubic box with periodic boundary conditions. Each of these systems has
been studied along three isochors ($\rho$ = 0.31, 0.32, and 0.33 for $\kappa$ = 3.0; $\rho$ = 0.25,
0.26, and 0.27 for $\kappa$ = 3.4; $\rho$ = 0.215, 0.225, and 0.235 for $\kappa$ = 3.8) at
several temperatures, starting from the high-temperature isotropic phase down to the nematic
phase across the I-N phase boundary. All quantities are given in reduced units defined in terms of
the Gay-Berne potential parameters $\epsilon_{0}$ and $\sigma_{0}$: length in units of $\sigma_{0}$,
temperature in units of $\frac {\epsilon_{0}}{k_{B}}$, and time in units of $(\frac {\sigma_{0}^{2}m}
{\epsilon_{0}})^{1/2}$, m being the mass of the ellipsoids of revolution. The mass as well
as the moment of inertia of each of the ellipsoids of revolution have been set equal to unity.
The intermolecular potential is truncated at a distance $r_{cut}$and shifted such
that $U(r_{ij} = r_{cut}) = 0$, $r_{ij}$ being the separation between two ellipsoids of
revolution i and j. The equations of motion have been integrated using the velocity-verlet
algorithm with integration time step $dt = 0.0015$ \cite{Ilnytskyi-CPM-2002}. Equilibration
has been done by periodic
rescaling of linear and angular velocities of particles. This has been done for a time period of
$t_{q}$ following which the system has been allowed to propagate with a constant energy for a
time period of $t_{e}$ in order to ensure equilibration upon observation of no drift of
temperature, pressure, and potential energy.  The data collection has been executed in a
microcanonical ensemble. At each state point, local potential energy minimization has been
executed by the conjugate gradient technique for a subset of $200$ statistically independent
configurations. The landscape analysis has been done with a system size of $256$ ellipsoids
of revolution, which is big enough for having no qualitative change in the results due to
the system size \cite{Chakrabarti-PNAS-2006}. Minimization has been performed with three
position coordinates and two Euler angles for each particle, the third Euler angle being
redundant for ellipsoids of revolution.

\section{Results and discussion}

The single-particle second rank orientational
time correlation function (OTCF) $C_{2}^{s}(t)$ is defined by
\noindent
\begin{equation}
C_{2}^{s}(t)=\frac{<\sum_{i}P_{2}({\bf \hat e}_{i}(t) \cdot
{\bf \hat e}_{i}(0)>}{<\sum_{i}P_{2}({\bf \hat e}_{i}(0) \cdot
{\bf \hat e}_{i}(0)>},
\end{equation}
\noindent
where $P_{2}$ is the second rank Legendre polynomial, ${\bf \hat e}_{i}$ is the
unit vector along the long axis of $i$th ellipsoid of revolution, and the angular
brackets stand for ensemble averaging.

\begin{figure}
\epsfig{file=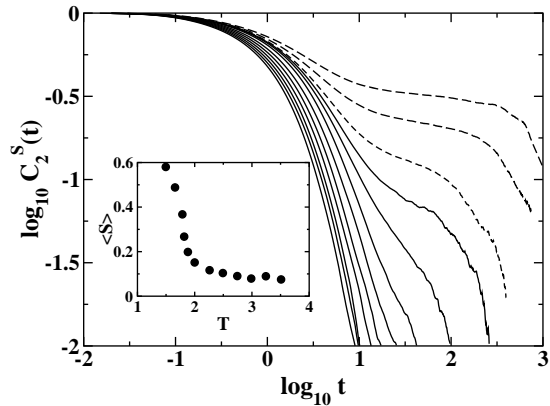,width=6cm,angle=270}
\caption{\label{fig:power} Time evolution of the single-particle second rank
orientational time correlation function $C_{2}^{s}(t)$ in a log-log plot for the
system with aspect ratio $\kappa = 3$. The time dependence is shown at several
temperatures across the isotropic-nematic transition along the isochor at density
$\rho = 0.33$. The solid lines denote the curves for the isotropic phase and the
dashed lines for the nematic phase. The inset shows the average orientational
order parameter $<S>$ as a function of temperature.}
\end{figure}

Figure \ref{fig:power} shows the time evolution of the single-particle second rank
OTCF for one of the three systems considered here as
the temperature is lowered along an isochor from the high-temperature isotropic
phase down to the nematic phase across the I-N phase boundary. In the inset, the
average orientational order parameter $<S>$ is shown as a function of temperature
along the isochor \cite{Zannoni-Book-2000}. The variation of $<S>$ with
temperature serves to locate the I-N phase boundary. In the present study,
the I-N transition temperature $T_{I-N}$ is taken as the temperature at which
$<S>$ of the system is $0.35$. For each aspect ratio, three isochors at different
densities have been considered. The qualitative behavior has been found to be the
same for all the three systems along all the isochors studied (data not shown).
The emergence of the power law decay in the isotropic phase near the I-N
transition is evident in all the cases as a universal characteristic of I-N 
transition \cite{Chakrabarti-PRL-2005}. As the I-N phase boundary is crossed upon
cooling, the advent of two power law decay regimes separated by an intervening
plateau at short-to-intermediate times imparts a step-like feature to the temporal
behavior of the second rank OTCF. Such a feature bears remarkable similarity to
what is observed for supercooled liquids as the glass transition is approached
from the above \cite{Kammerer-PRE-1997,DeMichele-PRE-2001}. While for the supercooled
liquid the emargence of step-like feature in the OTCF is well understood as a consequence of
$\beta$ relaxation, the origin of such a feature observed for liquid crystal defied of reliable explanation.

\begin{figure}
\epsfig{file=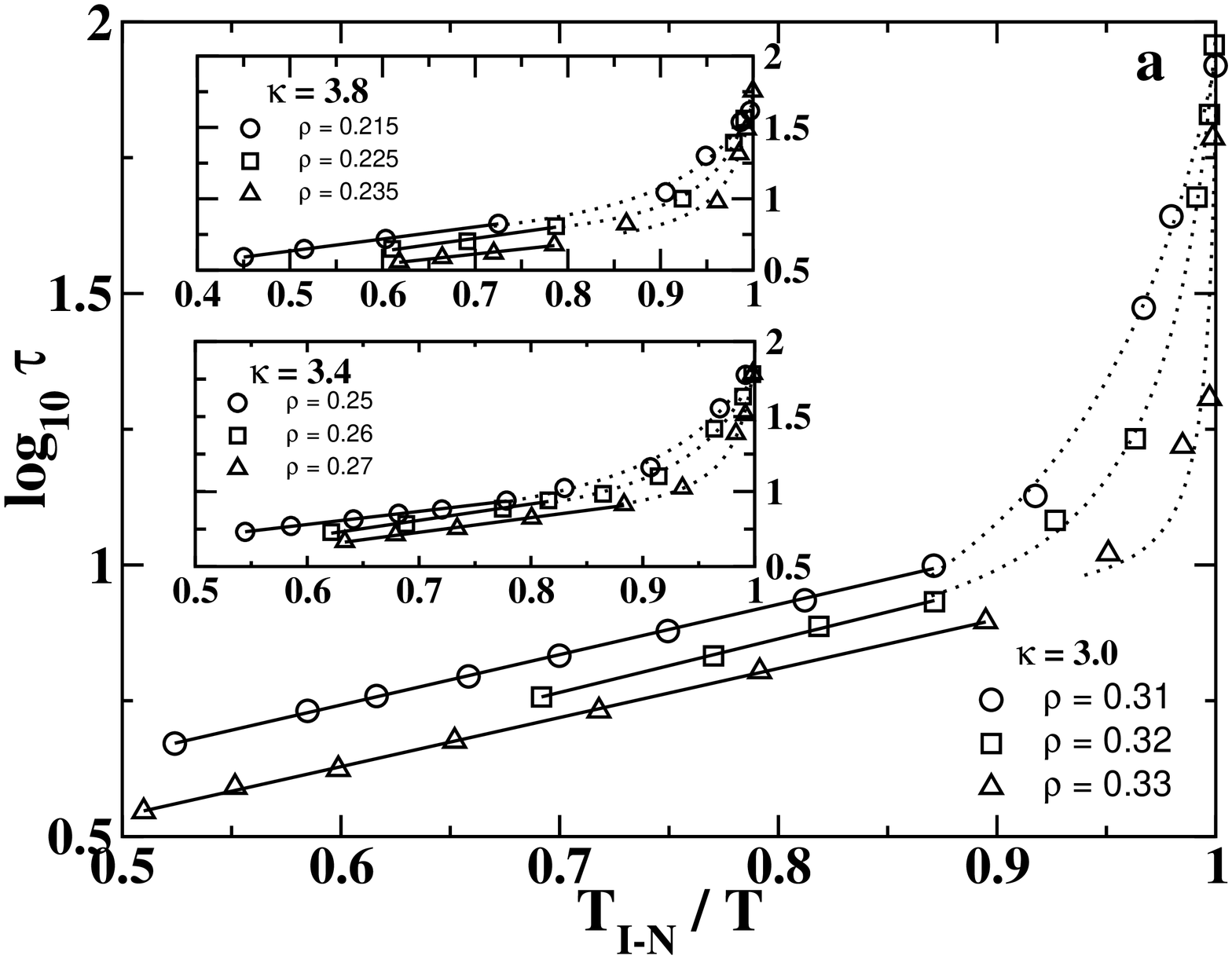,width=6cm,angle=270}
\epsfig{file=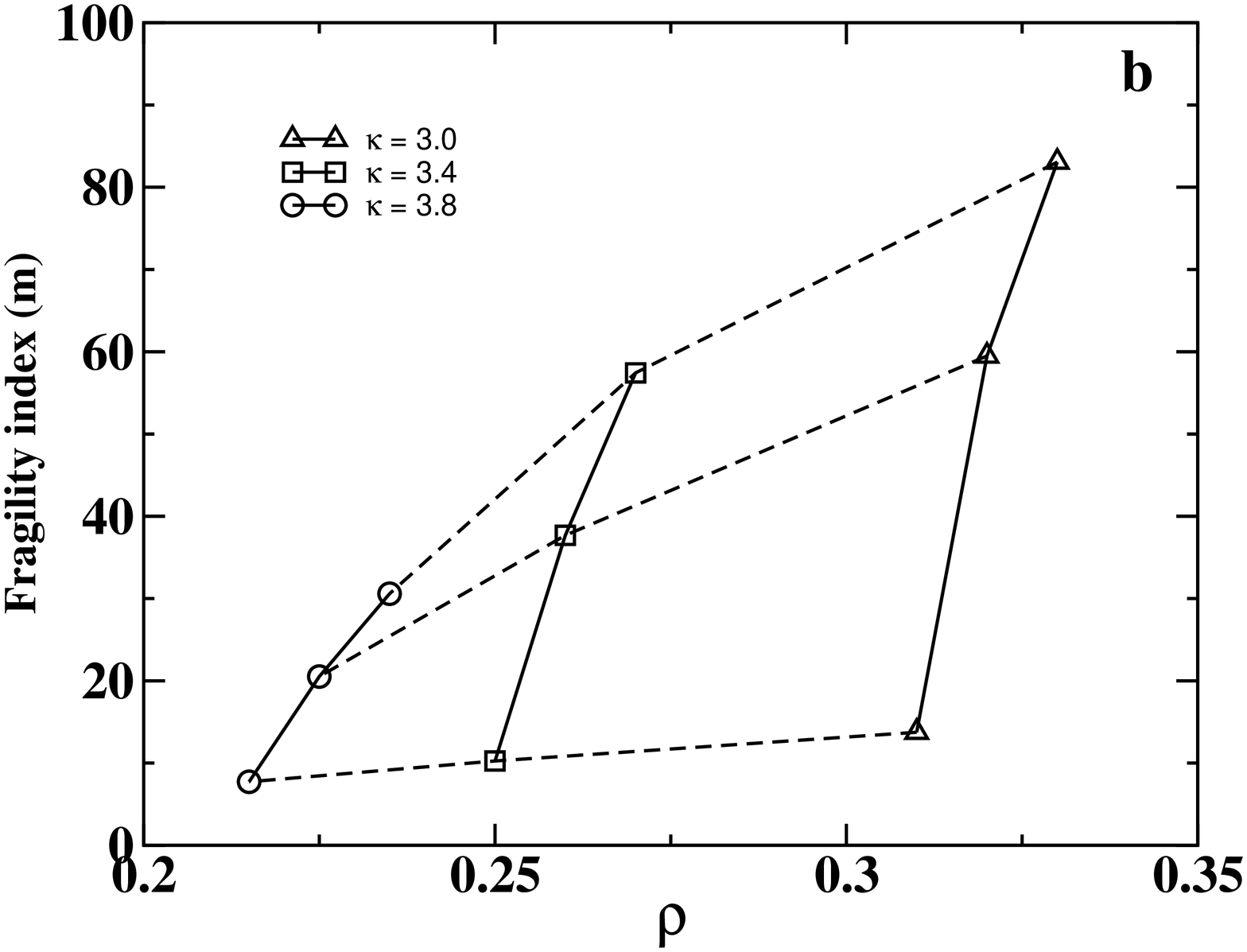,width=6cm,angle=270}
\caption{\label{fig:frgl}(a) The orientational correlation time $\tau$ in the
logarithmic scale as function of the inverse of the scaled temperature, the
scaling being done by the isotropic to nematic transition temperature $T_{I-N}$.
For the insets, the horizontal and the vertical axis labels read same as that of
the main frame and are thus omitted for clarity. Along each isochor, the solid
line is the Arrhenius fit to the subset of the high-temperature data and the
dotted line corresponds to the fit to the data near the isotropic-nematic phase
boundary with the VFT form. (b) The fragility index $m$ shown as a function of
density for different aspect ratios. The dashed lines are guide to the eye to
illustrate that the change in the fragility index for a given density difference
($\Delta\rho$) increases with the decrease in the aspect ratio.}
\end{figure}

We estimate the orientational correlation time $\tau$ as the time taken for
$C_{2}^{s}(t)$ to decay by $90\%$, i.e., $C_{2}^{s}(t=\tau) = 0.1$. Figure
\ref{fig:frgl}(a) shows $\tau$ in the
logarithmic scale as a function of the inverse temperature along the three
isochors for each of the three systems considered. We have scaled the temperature
by $T_{I-N}$ in the spirit of Angell's plot, that displays the shear viscosity (or
the structural relaxation time, the inverse diffusivity, etc.) of glass-forming
liquids as a function of the inverse of the scaled temperature, the scaling being
done in the latter case by the glass transition temperature $T_{g}$
\cite{Angell-JPCS-1988,Angell-JNCS-1991}. For all the three systems, two distinct
features are common: (i) in the isotropic phase far away from the I-N transition,
the orientational correlation time $\tau$ exhibits the Arrhenius
temperature dependence, i.e., $\tau(T) = \tau_{0}exp[E/(k_{B}T)]$, where the
activation energy $E$ and the pre-factor $\tau_{0}$ are both independent of
temperature; (ii) in the isotropic phase near the I-N transition, the temperature
dependence of $\tau$ shows marked deviation
from the Arrhenius behavior and can be well described by the Vogel-Fulcher-Tammann
(VFT) equation $\tau(T) = \tau_{0}exp[B/(T-T_{VFT})]$, where $\tau_{0}$, $B$, and
$T_{VFT}$ are constants, independent of temperature. Again these features bear
remarkable similarity with those observed for fragile glass-forming liquid. A
non-Arrhenius temperature behavior is taken to be the signature of fragile
liquids. For fragile liquids, the temperature dependence of the shear viscosity
follows the Arrhenius behavior far above $T_{g}$ and can be fitted to the VFT
functional form in the deeply supercooled regime near $T_{g}$
\cite{Angell-JPCS-1988,Angell-JNCS-1991}.

The striking resemblance in the dynamical behavior described above between the
isotropic phase of thermotropic liquid crystals near the I-N transition and
supercooled liquids near the glass transition has prompted us to attempt a
quantitative measure of glassy behavior near the I-N transition.  For supercolled liquids, 
one quantifies the dynamics by a parameter called fragility index which measures the rapidity 
at which the liquid's propeties (such as viscosity) change as the glassy state is approached. 
In the same spirit \cite{Bohmer-JCP-1993} that offers a quantitative estimation of the fragile
behavior of supercooled liquids, we here define the fragility index
$m$ of a thermotropic liquid crystalline system as
\noindent
\begin{equation}
m = \left. \frac{dlog \tau(T)}{d(T_{I-N}/T)} \right|_{T=T_{I-N}}.
\end{equation}
Figure \ref{fig:frgl}(b) shows the density dependence of the fragility index for
the three systems with different aspect ratios. For a given aspect ratio, the
fragility index increases with increasing density, the numerical values of the
fragility index $m$ being comparable to those of supercooled liquids. The change
in the fragility index for a given density difference ($\Delta\rho$) increases
with the decrease in the aspect ratio.

\begin{figure}
\epsfig{file=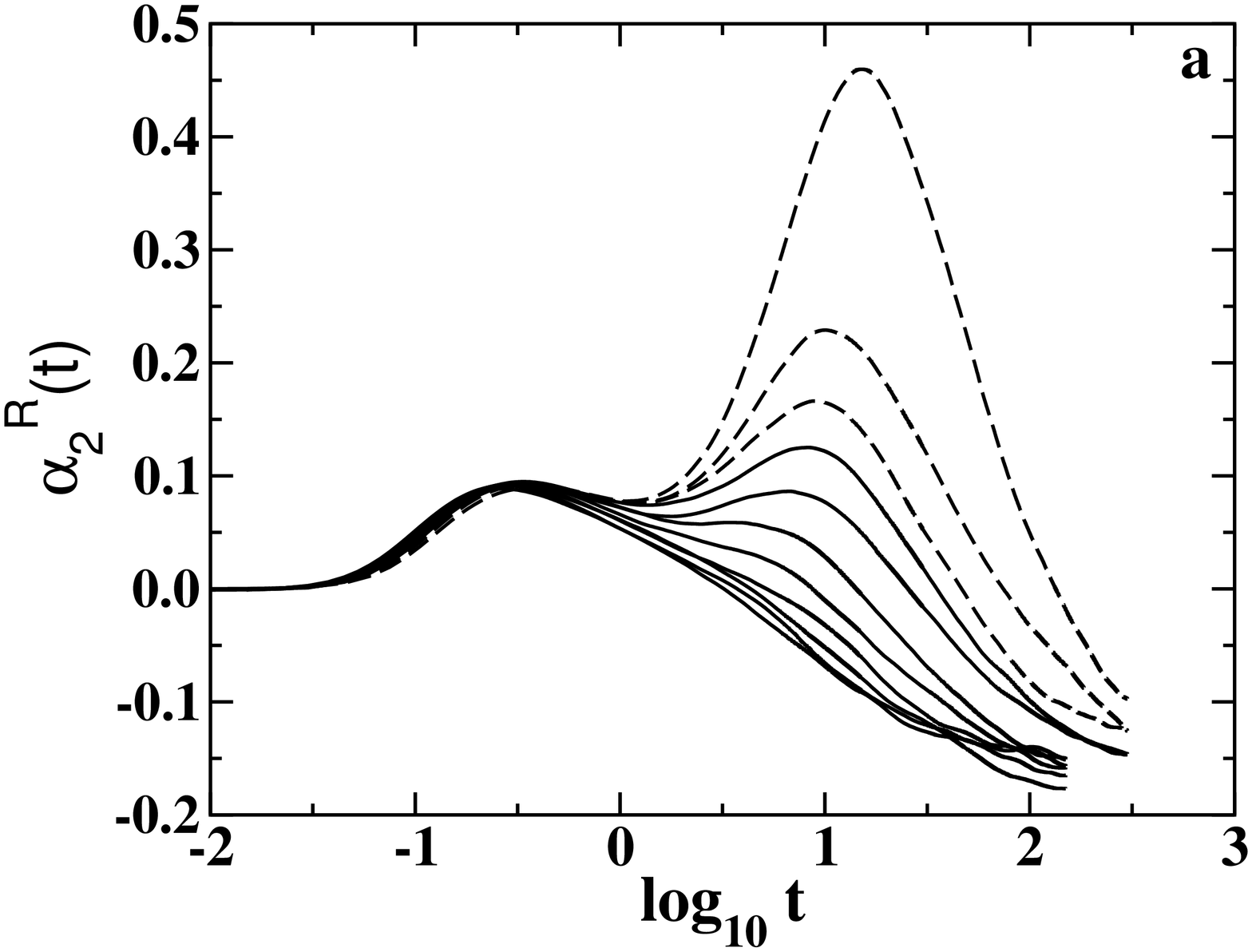,width=6cm,angle=270}
\epsfig{file=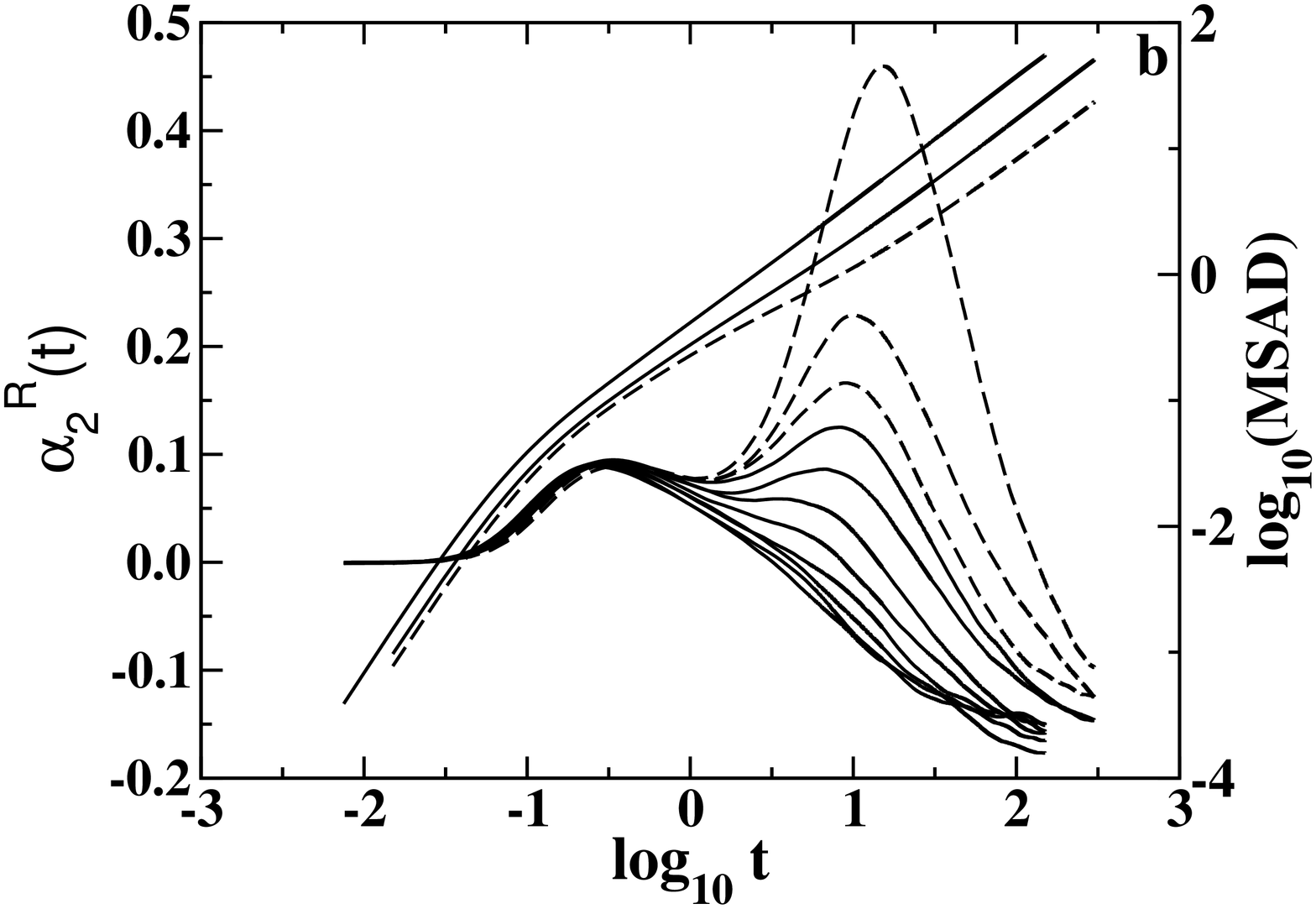,width=6cm,angle=270}
\epsfig{file=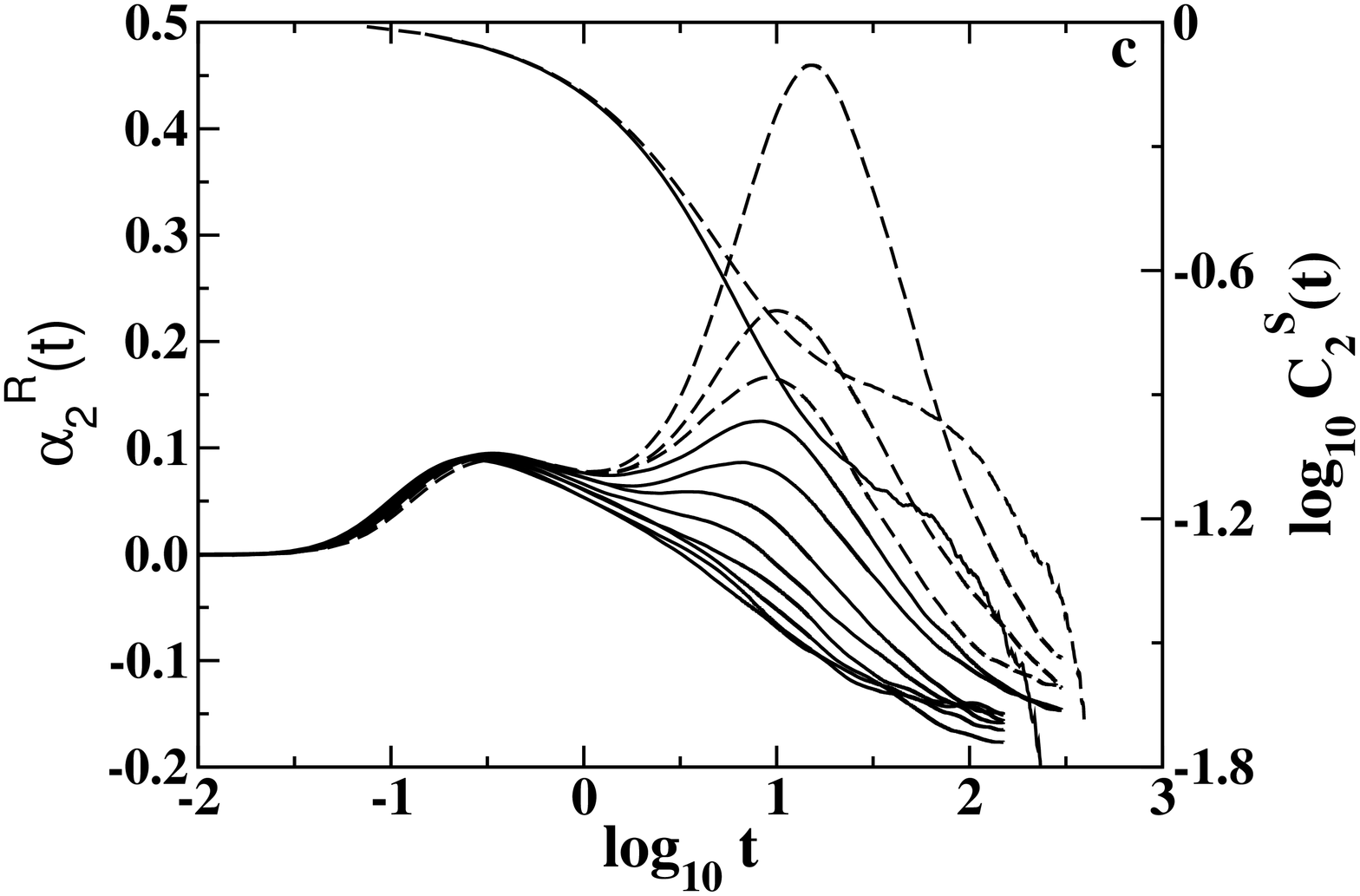,width=6cm,angle=270}
\caption{\label{fig:rngp}(a) Time evolution of the rotational non-Gaussian parameter
$\alpha_{2}^{R}(t)$ in a semi-log plot for the system with aspect ratio
$\kappa = 3$. The time dependence is shown at several temperatures across the
isotropic-nematic (I-N) transition along an isochor at density $\rho = 0.33$. (b) On a
different scale along the vertical axis (appearing on the right), time
evolution of the mean square angular deviation $<\Delta {\bm \phi}^{2}(t)>$ is shown in a
log-log plot for three temperatures: the highest temperature studied in the isotropic
phase and the other two temperatures that are nearest to the I-N transition in the from both side 
along with the time evolution of $\alpha_{2}^{R}(t)$, and (c) On a different scale along the 
vertical axis (appearing on the right), the time evolution of the single-particle second rank 
orientational time correlation function $C_{2}^{s}(t)$ is shown in a log-log plot for the two 
temperatures that are closest to the I-N transition on either side along with the time 
evolution of $\alpha_{2}^{R}(t)$.}
\end{figure}

Another hallmark of fragile glass-forming liquids is spatially heterogeneous
dynamics \cite{Ediger-ARPC-2000} reflected in non-Gaussian dynamical behavior
\cite{Shell-JPCM-2005}. It is intuitive that the growth of the pseudo-nematic
domains, characterized by local nematic order, in the isotropic phase near the I-N
transition would result in heterogeneous dynamics in liquid crystals. We have,
therefore, monitored the time evolution of the rotational non-Gaussian parameter
(NGP) \cite{Mazza-PRL-2006}, $\alpha_{2}^{R}(t)$, which in the present case is
defined as
\begin{equation}
\alpha_{2}^{R}(t)= \frac{<\Delta {\bm \phi}^{4}(t)>}{2 <\Delta {\bm \phi}^{2}(t)>
^{2}} - 1,
\end{equation}
where
\begin{equation}
<\Delta {\bm \phi}^{2n}(t)> = \frac{1}{N}\sum_{i=1}^{N} <|{\bm \phi}_{i}(t) -
{\bm \phi}_{i}(0)|^{2n}>.
\end{equation}
Here ${\bm \phi}_{i}$ is the rotation vector like the position vector ${\bf r}_{i}$
appears incase of translational NGP of
$i$th ellipsoid of revolution, the change of which is defined by
$\Delta {\bm \phi}_{i}(t) = {\bm \phi}_{i}(t) - {\bm \phi}_{i}(0) =
\int_{0}^{t}dt^{\prime} {\bm \omega}(t^{\prime})$,
$\omega_{i}$ being the corresponding angular velocity \cite{Kammerer-PRE-1997,
DeMichele-PRE-2001}, and $N$ is the number of ellipsoids of revolution in the
system. NGP will have value equal to zero when system dynamics is spatially homogeneous
and will have a non-zero value when the system dynamics is spatially heterogeneous.
As a typical behavior, Fig. \ref{fig:rngp}(a) shows the time dependence of
the rotational NGP for one of the systems at several temperatures across the I-N
transition along an isochor. On approaching the I-N transition upon cooling, a
bimodal feature starts appearing with the growth of a second peak, which
eventually becomes the dominant one, at longer times.  

We further investgate the apearance of this bimodal feature in NGP plot.  To this 
end we calculate mean square angular deviation (MSAD) of the system at different temperatures 
starting from high temperature isotropic phase to low temperature nematic phase. The appearance 
of the bimodal feature in the rotational NGP is accompanied by a signature of a sub-diffusive 
regime in the temporal evolution of the mean square angular deviation (MSAD), the time scale of 
the short-time peak and that of the onset of the sub-diffusive regime being comparable, as shown 
in  Fig. \ref{fig:rngp}(b). We further note that the dominant peak appears on a time scale which 
is comparable to that of onset of the diffusive motion in orientational degrees of freedom (ODOF) 
as evident in Fig. \ref{fig:rngp}(b) and similar feature has been observed recently for glassy 
systems \cite{Mazza-PRL-2006} also.

We also observe that the time scale at which long-time peak appears is also comparable to
the onset of the plateau that is observed in the time evolution of $C_{2}^{s}(t)$, as shown in
Fig. \ref{fig:rngp}(c).

\subsection{Mode coupling theory analysis} 

This striking similarities of the dynamics between liquid crystals near I-N transition and supercooled 
liquid near glass transition are also supported by the use of Mode Coupling Theory (MCT) to explain the 
dynamics of both the systems.  While MCT was used first for the supercooled liquid, recently it has been 
used for liquid crystals also.  MCT theory devloped by Gottke et al. \cite{Gottke-JCP-2002} predicts that 
near I-N transition, the low frequency rotational memory kernel should diverge in a power law fashion.
\begin{equation}
 M_{R}(z) \approx \frac {A}{z^{\alpha}} 
\end{equation}
Mean field treatment gives $\alpha = 0.5$. Invoking the rank ($l$) dependence of the memory function, 
the single particle OTCF can be written as \cite{Ravichandran-IRPC-1995,Bagchi-ACP-1991,
Hubbard-JCP-1978,Bagchi-JML-1998}
\begin{equation}
C_{l}^{s}(z) =  \left [z + \frac {l(l+1)k_{B}T}{I[z+\Gamma _{l}(z)}\right]^{-1}
\end{equation}
The above equation can be Laplace inverted to obtain a short-to-intermadiate power law decay in 
$C_{2}^{s}(t)$  which is a universal characteristic of the I-N transition for several model liquid 
crystals \cite{Jose-PRE-2005,Chakrabarty-PRE-2006,Chakrabarti-PRL-2005}. Recently, Li et al. 
\cite{Li-JCP-2006}  have showed that it is also possible to formulate a schematic model that combines 
short-to-intermediate time relaxation with long time relaxation. In their model, they have expressed the 
total memory functon ($M(t)$) as the sum of mode coupling memory function ($M_{MCT}(t)$) and Landau-de 
Gennes memory function ($M_{LdG}(t)$).
\begin{equation}
M(t) = M_{MCT}(t) + M_{LdG}(t)
\end{equation}
where
\begin{equation}
M_{MCT}(t) = \Omega ^{2}K(t)
\end{equation}
$\Omega$ is the characteristic frequecny and $K(0) = 1$. Time dependence of $K(t)$ can be written 
expressed in terms of the memory function $m(t)$ \cite{Li-JCP-2006} and $m(t)$ has the following form
\begin{equation}
m(t) = \kappa \phi (t) \phi _{1} (t)  
\end{equation}
Where $\phi (t)$ is the autocorrelation function of the anisotropy of the polarizability and 
$\phi _{1}(t)$ is the solution of a $F_{12}$ schematic model for what is referred to as the 
density correlator. $\kappa$ being the coupling constant between them. Now, $M_{LdG}(t)$ can be 
written as
\begin{equation}
M_{LdG}(t) = \Gamma \delta (t)
\end{equation}
Here $\Gamma ^{-1}$  is the relaxation time ($\Gamma ^{-1} = \tau_{LdG}$) and it diverges as 
$(T-T^{\star})^{-1}$ as the critical temperature $T^{\star}$ of the I-N transition is approached 
from the above. Following the calculation of  Ref. \cite{Li-JCP-2006} , one can get two important 
relaxation equattions. 
\begin{equation}
\ddot{\phi_{1}}(t) = -\Omega_{1}^{2} \phi_{1}(t) - \mu_{1}\dot{\phi_{1}}(t) -  
\Omega_{1}^{2}\int_{0}^{t} dt^{\prime} m_{1}(t-t^{\prime})\dot{\phi_{1}}(t^{\prime}) 
\end{equation}  
With the initial conditions $\phi_{1}(0) = 1$ and $\dot{\phi_{1}}(0) = 0$ 
and
\begin{equation}
\ddot{\phi} (t) = - (\Omega^{2} + \mu \Gamma ) \phi (t) - (\mu + \Gamma) \dot{\phi} (t) - 
\Omega^{2} \int_{0}^{t} dt^{\prime} m(t-t^{\prime}) \dot{\phi} (t^{\prime})  - 
\Omega^{2}\Gamma \int_{0}^{t} dt^{\prime} m(t-t^{\prime}) \phi (t^{\prime})
\end{equation}
with the initial conditions $\phi (0) = 1$ and $\dot{\phi}(0) = - \Gamma$ . $\mu_{1}$ and $\mu$ are 
the damping constants. Eq. $17$ is identical to one gets from the MCT analysis of the supercooled liquids. 
Difference between this scematic model and one applied for the supercooled liquid is in Eq. $18$ . If 
$\Gamma$ is set equal to $0$ in Eq. $18$ , the supercooled model is recovered. Eq. $18$ is  the 
orientational correlation function coupled to the density correlation function with specific new terms 
that account for the long-time portion of the relaxation profile that has been previously described by 
LdG theory.

\subsection{Energy landscape analysis}

\begin{figure}
\epsfig{file=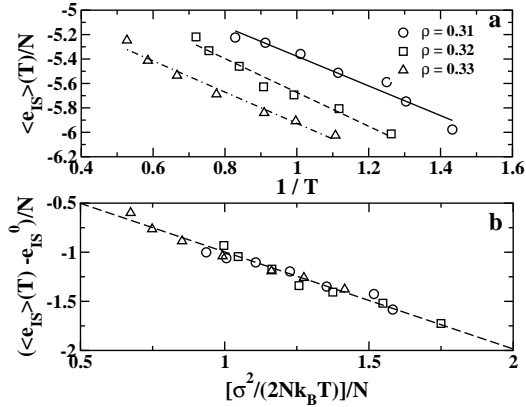,width=6cm,angle=270}
\caption{\label{fig:lscp} The energy landscapes as explored by the system with the
aspect ratio $3$ on variation of temperature along three isochors. (a) The average
inherent structure (IS) energy per particle as a function of the inverse
temperature at different densities. The solid line, dashed line, and dot-dashed
lines are the linear fits to the data at the densities $\rho = 0.31, 0.32, 0.33$,
respectively. (b) The displaced average IS energy per particle versus a scaled
inverse temperature along the same three isochors. If the Gaussian model for the
number density of the IS energy with a given energy is validated, a collapse of
the data for all densities is expected onto a straight line with negative unit
slope, that is drawn. The data are shown over the temperature regimes in which the
average IS energy is on a decline.}
\end{figure}

Several studies have attempted to interpret the fragility of glass-forming
liquids in terms of the features of the underlying energy landscapes
\cite{Wales-Book-2003,Sastry-Nature-1998,Scala-Nature-2000,Sastry-Nature-2001,
Martinez-Nature-2001,Salka-Nature-2001}. Energy landscape analysis gives the potential energy,
which devoids of any kind of thermal motions, of inherent structures of the parent
liquid and hence provide a better understanding of the structure and dynamics of the
paprent liquid.  A recent study on thermotropic liquid
crystals has reported the temperature dependent exploration of the energy
landscapes of a family of the Gay-Berne model systems across the mesophases
\cite{Chakrabarti-PNAS-2006}. The average inherent structure (IS) energy $<e_{IS}>$
has been found to fall as $<S>$ grows across the I-N phase boundary and through
the nematic phase in contrast to its insensitivity to the temperature in the
high-temperature isotropic phase and the low-temperature smectic-B phase
\cite{Chakrabarti-PNAS-2006}. Such a fall in the average IS energy is consistent
with a Gaussian form for the number density of inherent structures with energy
$e_{IS}$, that predicts a linear variation of $<e_{IS}>$ with the inverse temperature:
$<e_{IS}>(T)=e_{IS}^{0}-\sigma^{2}/2Nk_{B}T$, where $e_{IS}^{0}$ and
$\sigma$ are parameters independent of temperature and $k_{B}$ is the Boltzmann
constant \cite{Sastry-Nature-2001}. Note that this has been observed for a glassy
system \cite{Sastry-Nature-2001}, where the average IS energy also falls over a
temperature range \cite{Sastry-Nature-1998}. In Fig. \ref{fig:lscp}(a), we
demonstrate with the original and the most studied parameterization for the GB
pair potential GB$(3, 5, 2, 1)$ that the prediction holds good over the
temperature range where $<e_{IS}>$ is on a decline along all the three isochors
studied. It then follows that a plot of $<e_{IS}>(T) - e_{IS}^{0}$
versus $\sigma^{2}$/$2Nk_{B}T$ would result in a collapse of the $<e_{IS}>$ data for
all densities onto a straight line with negative unit slope. This is indeed found
to be true, as shown in Fig. \ref{fig:lscp}(b), implying the validity of the
Gaussian model in this case as well. It may be noted that when the distribution of
IS energy is Gaussian, the fragility of glass-forming liquids has been shown to
depend on the total number of inherent structures, the width of the Gaussian, and
the variation of the basin shape with the average IS energy
\cite{Sastry-Nature-2001}.

\section{Conclusion}

The origin of the glassy orientational dynamics, both single
particle and collective, of nematogens near the isotropic-nematic
transition has been addressed in several publications in recent
years \cite{Gottke-JCP-2002,Cang-JCP-2003,Jose-PRE-2005,Chakrabarti-PNAS-2006}. 
In these studies, the similarity between dynamics of supercooled
glassy liquid and liquid crystals has  been discussed in detail, but
no quantitative measure of the similarity was provided. The
fragility index introduced here serves to remove that lacuna. It is
indeed surprising that even the values of the fragility parameter are in
the range observed for glassy liquids as well. This is in agreement
with the repeated observation by Fayer and coworkers that the values
of the power law exponents observed in the two systems are quite
similar. Further understanding of the relaxation mechanism has been obtained
from a closure look on the heterogeneous dynamics. As figure 3 demonstrates, the
rotational non-Gaussian parameter shows a dramatic enhancement of hetrogeneous 
dynamics as the I-N phase boundary is approached. Unlike what is
found near the gas-liquid critical point \cite{Kunter-PRB-1982}, the
single-particle dynamics near the I-N phase boundary are observed to
be strongly affected by the approaching thermodynamic singularity.
We have discussed mode coupling theory approaches introduced to understand
anomalous dynamics observed in this problem.

\begin{center}
{\large \bf Acknowledgments}
It is a pleasure to thank Professor S. Sastry for helpful discussions. This work was
supported in parts by grants from DST and CSIR, India. B.J. acknowledges CSIR, India and D.C. acknowledges UGC, India for providing Research Fellowship.
\end{center}


\begin{references}

\bibitem{deGennes-book-1993} P. G. de Gennes and J. Prost, {\it The Physics of
Liquid Crystals} (Clarendon Press, Oxford, 1993).

\bibitem{Chandrasekhar-book-1992} S. Chandrasekhar, {\it Liquid Crystals}
(Cambridge University Press, Cambridge, 1992).

\bibitem{Gottke-JCP-2002} S. D. Gottke, H. Cang, B. Bagchi, and M. D. Fayer, J.
Chem. Phys. {\bf 116}, 6339 (2002).

\bibitem{Cang-JCP-2003} H. Cang, J. Li, V. N. Novikov, and M. D. Fayer, J. Chem.
Phys. {\bf 118}, 9303 (2003).

\bibitem{Jose-PRE-2005} P. P. Jose, D. Chakrabarti, and B. Bagchi, Phys. Rev. E
{\bf 71}, 030701(R) (2005).

\bibitem{Chakrabarty-PRE-2006} S. Chakrabarty, D. Chakrabarti, and B. Bagchi,
Phys. Rev. E. {\bf 73}, 061706 (2006).

\bibitem{Gay-JCP-1981} J. G. Gay and B. J. Berne, J. Chem. Phys. {\bf 74}, 3316
(1981).

\bibitem{Bates-JCP-1999} M. A. Bates and G. R. Luckhurst, J. Chem. Phys.
{\bf 110}, 7087 (1999).

\bibitem{Ilnytskyi-CPM-2002} J. M. Ilnytskyi and M. R. Wilson, Comput. Phys.
Commun. {\bf 148}, 43 (2002).

\bibitem{Chakrabarti-PNAS-2006} D. Chakrabarti and B. Bagchi, Proc. Natl. Acad.
Sci. USA   {\bf 103}, 7217 (2006).

\bibitem{Zannoni-Book-2000} C. Zannoni in {\it Advances in the Computer
Simulations of Liquid Crystals} (eds. P. Pasini and C. Zannoni) (Kluwer Academic
Publishers, Dordrecht, 2000).

\bibitem{Chakrabarti-PRL-2005} D. Chakrabarti, P. P. Jose, S. Chakrabarty and B. 
Bagchi, Phys. Rev. Lett. {\bf 95}, 197801 (2005).

\bibitem{Kammerer-PRE-1997}  S. K\"{a}mmerer, W. Kob, and R. Schilling, Phys. Rev.
E {\bf 56}, 5450 (1997).

\bibitem{DeMichele-PRE-2001} C. De Michele and D. Leporini, Phys. Rev. E
{\bf 63}, 036702 (2001).

\bibitem{Angell-JPCS-1988} C. A. Angell, J. Phys. Chem. Solids. {\bf 49}, 863
(1988).

\bibitem{Angell-JNCS-1991} C. A. Angell, J. Non-Cryst. Solids. {\bf 131-133}, 13
(1991).

\bibitem{Bohmer-JCP-1993} R. B\"{o}hmer, K. L. Ngai, C. A. Angell, and D. J.
Plazek, J. Chem. Phys. {\bf 99}, 4201 (1993).

\bibitem{Ediger-ARPC-2000} M. D. Ediger, Ann. Rev. Phys. Chem. {\bf 51}, 99
(2000).

\bibitem{Shell-JPCM-2005} M. S. Shell, P. G. Debenedetti, and F. H. Stillinger,
J. Phys.: Condens. Matter {\bf 17}, S4035 (2005).

\bibitem{Mazza-PRL-2006} M. G. Mazza, N. Giovambattista, F. W. Starr, and H. E.
Stanley, Phys. Rev. Lett. {\bf 96}, 057803 (2006).

\bibitem{Ravichandran-IRPC-1995} S. Ravichandran,  and B. Bagchi, Int. Rev. Phys. Chem. {\bf 14}, 271 (1995)

\bibitem{Bagchi-ACP-1991} B. Bagchi, and A. Chandra, Adv. Chem. Phys.  {\bf 80}, 1 (1991)

\bibitem{Hubbard-JCP-1978} J. B. Hubbard, and P. G. Wolynes, J. Chem. Phys. {\bf 69}, 998 (1978)

\bibitem{Bagchi-JML-1998} B. Bagchi, J. Mol. Liq. {\bf 77}, 177 (1998) 

\bibitem{Li-JCP-2006} J. Li, H. Cang, H. C. Andersen, and M. D. Fayer, J. Chem. Phys. {\bf 124}, 014902 (2006).  

\bibitem{Wales-Book-2003} D. J. Wales, {\it Energy Landscapes} (Cambridge
University Press, Cambridge, 2003).

\bibitem{Sastry-Nature-1998} S. Sastry, P. G. Debenedetti, and F. H. Stillinger,
Nature {\bf 393}, 554 (1998).

\bibitem{Scala-Nature-2000} A. Scala, F. W. Starr, E. L. Nave, F. Sciortino, and
H. E. Stanley, Nature {\bf 406}, 166 (2000).

\bibitem{Sastry-Nature-2001} S. Sastry, Nature {\bf 409}, 164 (2001).

\bibitem{Martinez-Nature-2001} L. -M. Martinez and C. A. Angell, Nature {\bf 410},
633 (2001).

\bibitem{Salka-Nature-2001} I. Salka-Voivod, P. H. Poole, and F. Sciortino, Nature
{\bf 412}, 514 (2001).

\bibitem{Kunter-PRB-1982} R. Kutner, K. Binder, and K. W. Kehr, Phys. Rev. B
{\bf 26}, 2967 (1982).

\end{references}
\end{document}